\documentclass[twoside,twocolumn,9pt]{article}
\usepackage{extsizes}
\usepackage[super,sort&compress,comma]{natbib} 
\usepackage[version=3]{mhchem}
\usepackage[left=1.5cm, right=1.5cm, top=1.785cm, bottom=2.0cm]{geometry}
\usepackage{balance}
\usepackage{mathptmx}
\usepackage{sectsty}
\usepackage{graphicx} 
\usepackage{lastpage}
\usepackage[format=plain,justification=justified,singlelinecheck=false,font={stretch=1.125,small,sf},labelfont=bf,labelsep=space]{caption}
\usepackage{float}
\usepackage{fancyhdr}
\usepackage{fnpos}
\usepackage[english]{babel}
\addto{\captionsenglish}{%
  
}
\usepackage{array}
\usepackage{droidsans}
\usepackage{charter}
\usepackage[T1]{fontenc}
\usepackage[usenames,dvipsnames]{xcolor}
\usepackage{setspace}
\usepackage[compact]{titlesec}
\usepackage{hyperref}

\usepackage{bm}

\usepackage{soul}
\usepackage[usenames,dvipsnames]{xcolor}
\usepackage{color}
\usepackage{placeins}

\definecolor{cream}{RGB}{222,217,201}

\begin{document}

\pagestyle{fancy}
\thispagestyle{plain}
\fancypagestyle{plain}{
\renewcommand{\headrulewidth}{0pt}
}

\makeFNbottom
\makeatletter
\renewcommand\LARGE{\@setfontsize\LARGE{15pt}{17}}
\renewcommand\Large{\@setfontsize\Large{12pt}{14}}
\renewcommand\large{\@setfontsize\large{10pt}{12}}
\renewcommand\footnotesize{\@setfontsize\footnotesize{7pt}{10}}
\makeatother

\renewcommand{\thefootnote}{\fnsymbol{footnote}}
\renewcommand\footnoterule{\vspace*{1pt}%
\color{cream}\hrule width 3.5in height 0.4pt \color{black}\vspace*{5pt}} 
\setcounter{secnumdepth}{5}

\makeatletter 
\renewcommand\@biblabel[1]{#1}            
\renewcommand\@makefntext[1]%
{\noindent\makebox[0pt][r]{\@thefnmark\,}#1}
\makeatother 
\renewcommand{\figurename}{\small{Fig.}~}
\sectionfont{\sffamily\Large}
\subsectionfont{\normalsize}
\subsubsectionfont{\bf}
\setstretch{1.125} 
\setlength{\skip\footins}{0.8cm}
\setlength{\footnotesep}{0.25cm}
\setlength{\jot}{10pt}
\titlespacing*{\section}{0pt}{4pt}{4pt}
\titlespacing*{\subsection}{0pt}{15pt}{1pt}

\fancyfoot{}
\fancyfoot[LO,RE]{\vspace{-7.1pt}\includegraphics[height=9pt]{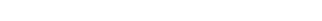}}
\fancyfoot[CO]{\vspace{-7.1pt}\hspace{13.2cm}\includegraphics{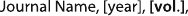}}
\fancyfoot[CE]{\vspace{-7.2pt}\hspace{-14.2cm}\includegraphics{head_foot/RF}}
\fancyfoot[RO]{\footnotesize{\sffamily{1--\pageref{LastPage} ~\textbar  \hspace{2pt}\thepage}}}
\fancyfoot[LE]{\footnotesize{\sffamily{\thepage~\textbar\hspace{3.45cm} 1--\pageref{LastPage}}}}
\fancyhead{}
\renewcommand{\headrulewidth}{0pt} 
\renewcommand{\footrulewidth}{0pt}
\setlength{\arrayrulewidth}{1pt}
\setlength{\columnsep}{6.5mm}
\setlength\bibsep{1pt}

\makeatletter 
\newlength{\figrulesep} 
\setlength{\figrulesep}{0.5\textfloatsep} 

\newcommand{\topfigrule}{\vspace*{-1pt}%
\noindent{\color{cream}\rule[-\figrulesep]{\columnwidth}{1.5pt}} }

\newcommand{\botfigrule}{\vspace*{-2pt}%
\noindent{\color{cream}\rule[\figrulesep]{\columnwidth}{1.5pt}} }

\newcommand{\dblfigrule}{\vspace*{-1pt}%
\noindent{\color{cream}\rule[-\figrulesep]{\textwidth}{1.5pt}} }

\makeatother

\twocolumn[
  \begin{@twocolumnfalse}
{\includegraphics[height=30pt]{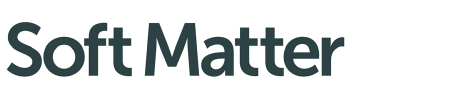}\hfill\raisebox{0pt}[0pt][0pt]{\includegraphics[height=55pt]{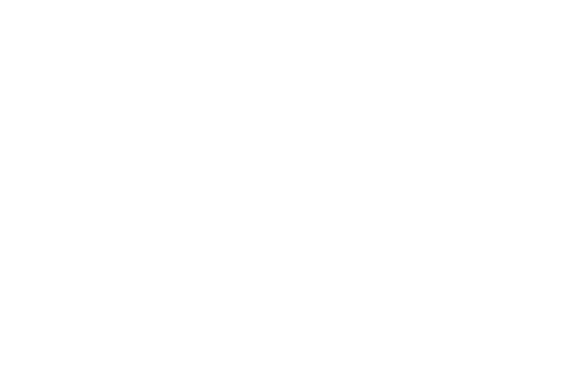}}\\[1ex]
\includegraphics[width=18.5cm]{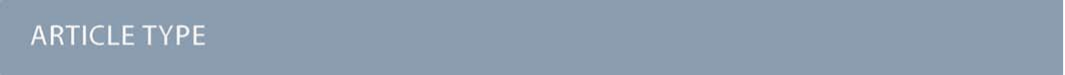}}\par
\vspace{1em}
\sffamily
\begin{tabular}{m{4.5cm} p{13.5cm} }

\includegraphics{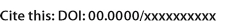} & \noindent\LARGE{\textbf{Elasticity tunes mechanical stress localization around active topological defects$^\dag$}} \\
\vspace{0.3cm} & \vspace{0.3cm} \\

 & \noindent\large{Lasse Bonn\textit{$^{a}$}, Aleksandra Arda\v{s}eva$^{a}$ and Amin Doostmohammadi,\textit{$^{a\ddag}$}} \\

\includegraphics{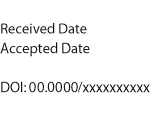} & \noindent\normalsize{
Mechanical stresses are increasingly found to be associated with various biological functionalities.
At the same time, topological defects are being identified across a diverse range of biological systems and are points of localized mechanical stress.
It is therefore important to ask how mechanical stress localization around topological defects is controlled.  
Here, we use continuum simulations of nonequilibrium, fluctuating and active nematics to explore the patterns of stress localization, as well as their extent and intensity around topological defects.
We find that by increasing the orientational elasticity of the material, the isotropic stress pattern around topological defects is changed substantially from a stress dipole characterized by symmetric compression-tension regions around the core of the defect to a localized stress monopole at the defect position.
Moreover, we show that elastic anisotropy alters the extent and intensity of the stresses, and can result in the dominance of tension or compression around defects.
Finally, including both nonequilibrium fluctuations and active stress generation, we find that the elastic constant tunes the relative effect of each, leading to the flipping of tension and compression regions around topological defects.
This flipping of the tension-compression regions only by changing the elastic constant presents an interesting, simple, way of switching the dynamic behavior in active matter by changing a passive material property.
We expect these findings to motivate further exploration tuning stresses in active biological materials by varying material properties of the constituent units.} \\

\end{tabular}

 \end{@twocolumnfalse} \vspace{0.6cm}

  ]

\renewcommand*\rmdefault{bch}\normalfont\upshape
\rmfamily
\section*{}
\vspace{-1cm}


\footnotetext{\textit{$^{a}$~Niels Bohr Institute, University of Copenhagen, Blegdamsvej 17, Copenhagen, Denmark}}

\footnotetext{\ddag~E-mail: doostmohammadi@nbi.ku.dk}



\section{Introduction}

Mechanics have been shown to play an indispensable role in governing cell behavior, from differentiation of stem cells~\cite{hayward2021tissue, vining2017mechanical}, to the migration of epithelial cells~\cite{ladoux_mechanobiology_2017}, and the establishment of bacterial size and shape~\cite{persat2015mechanical}. 
Mechanical stimuli and cell response are connected through the process of {\it mechanotransduction}, which translates physical forces into biochemical responses in eukaryotes and prokaryotes~\cite{wang2017review, dufrene2020mechanomicrobiology}.
Such biochemical responses, in turn, activate signaling pathways leading to various feedbacks to the cell behavior, controlling vital cell functions such as migration, proliferation, and cell death~\cite{pocaterra_yaptaz_2020, panciera2017mechanobiology}.
Mechanotransduction plays a role at subcellular scales, controlling the behavior of the cell cytoskeleton~\cite{mathieu2019intracellular}, and at the tissue scale impacting collective cell behaviors such as jamming~\cite{boocock2023interplay}, wound healing~\cite{brugues2014forces} and cell extrusion~\cite{ladoux_mechanobiology_2017,monfared2023mechanical}.
As such, identifying regions of mechanical stress localization within living materials is of considerable importance in various biophysical applications, as these localized stress areas present hot spots of signaling activation in a wide range of living materials.

Recently, through the analogy between several distinct living materials and liquid crystals, topological defects have emerged as regions of stress localization with potential biological functions~\cite{saw_topological_2017}.
Topological defects -- singularities in the orientation field -- have been found to be at the core of several important biological functions~\cite{doostmohammadi_physics_2021}, such as cell death and extrusion~\cite{saw_topological_2017}, mound formation in stem cells~\cite{kawaguchi_topological_2017}, bacterial competition~\cite{meacock_bacteria_2021}, limb origination in morphogenesis~\cite{maroudas-sacks_topological_2021}, and differentiation of myoblasts~\cite{guillamat2022integer}, among others.

In two-dimensional nematic systems, topological defects are generated as pairs of $\pm1/2$ charged defects~\cite{gennes_physics_1993}. 
As regions of high deformation, defects allow for local phenomena, not possible in the strongly aligned nematic phase.
In active systems, while the trefoil-shaped $-1/2$ defect moves diffusively, the $+1/2$ has a polar, comet shape, which allows the defect to move along its head-tail axis~\cite{doostmohammadi_active_2018}, where the `head' consists mostly of bend deformation and the `tail' consists mostly of splay deformation. 
The isotropic stress of the  $+1/2$ defect has a distinctive dipole pattern~\cite{saw_topological_2017}, generating regions of compression and tension around the defect core.
Therefore, the defect -- a feature of the orientation field of the cells -- by way of generating flow and stress features, affects the mechanics of the cell layer in a highly local way and becomes an actor in the mechanotransduction picture of tissue biology.

This is most readily demonstrated in the case of cell extrusion in mammalian epithelia (MDCK cells)~\cite{saw_topological_2017}, where $+1/2$ topological defects with a dipolar tension-compression stress pattern were shown to cause extrusion of cells from the layer.
Likewise, tension at the center of a $-1/2$ defect was shown to lead to hole formation in MDCK monolayers~\cite{sonam_2023_mechanical}.
More generally, in the tissue bulk, compression has been widely linked to cell removal, while tension has been associated with cell division~\cite{zulueta-coarasa_role_2022}.
While significant focus has been put on identifying topological defects and their potential roles across biological matter, much less is known about what controls the localized stress patterns around the defects. In particular, to connect the topological defects to biological functionality, it is of considerable importance to understand how the intensity of the stress and its spread around defects are affected by the intrinsic properties of the living material such as its elasticity.
A study of the stresses generated around topological defects is therefore in order.

In passive liquid crystals, orientational elasticity -- the resistance of the material to orientational deformations -- is known to play a central role in the generation of both flow and stress of topological defects~\cite{toth_hydrodynamics_2002, khoromskaia2017vortex, bonn_fluctuation-induced_2022}.
Moreover, in most experimental realizations of passive liquid crystals, the elastic response to orientational deformations is shown to be characterized by two distinct elastic constants that penalize bend and splay deformation modes~\cite{gennes_physics_1993}.
The difference in splay and bend elastic constants is especially relevant around $+1/2$ topological defects because the defect combines a bend region at the head and a splay region at the tail.
Indeed, experiments on actin filament-myosin motor protein mixtures have shown that increasing the bend elastic constant by adding microtubules results in alteration of the $+1/2$ defect shape from a rounder \rotatebox[origin=c]{180}{U} shape for a high splay elastic constant to a pointed $\Lambda$ shape for a high bend elastic constant~\cite{zhang_interplay_2018, kumar_tunable_2018}.
A similar change in defect shape has been observed in continuum and particle-based simulations~\cite{kumar_tunable_2018,joshi_interplay_2019}.
Nevertheless, the majority of theoretical and modeling studies have been limited to using the single elastic constant approximation~\cite{thampi_instabilities_2014, khoromskaia2017vortex, doostmohammadi_active_2018,khoromskaia_active_2023} and important questions about the role of elasticity on governing the mechanical stresses around topological defects remain unanswered.

Here, we focus on addressing these questions using a continuum model of a nematic liquid crystal. 
In particular, we explore the impact of orientational elasticity on patterns of stress localization around topological defects in the presence of nonequilibrium fluctuations and active stresses.
In both cases, we find a significant increase in the intensity and spread of mechanical stress around defects upon increasing elasticity and show an intriguing crossover from a stress dipole characterized by compression-tension regions around the defects, to a localized stress monopole at the defect position.
Furthermore, going beyond the single elastic constant approximation, we find that different bend and splay elastic constants can result in a substantial alteration of stress patterns around topological defects leading to a switch from intense tension and diffuse compression to intense compression and diffuse tension.
Combining both active fluctuations and active stresses, we find that the elasticity can tune the relative effect of each, making it possible to invert the stress pattern with elasticity alone.

The paper is organized as follows: first we review the continuum nematohydrodynamics equations and the simulation method (section~\ref{sec:meth}).
This is then followed by the results of varying the single elastic constant for a fluctuating nematics in section~\ref{sec:singleK}.
We then present results of varying bend and splay elasticities in section~\ref{sec:bendsplay}. 
Finally, we provide characterization of elasticity effects in the presence of active stresses in section~\ref{sec:active}, show the effect of varying the
single elastic constant on -1/2 defects in section~\ref{sec:neg}, and discuss the relevance of our findings and potential experimental tests of our predictions in section~\ref{sec:disc}.

\section{Methods\label{sec:meth}}

We implement fluctuating two-dimensional nematohydrodynamics as described in \cite{bonn_fluctuation-induced_2022}.
The equations are implemented with a hybrid lattice-Boltzmann approach~\cite{Marenduzzo_hlb_2007}.

The nematic order parameter is described by a tensor ${Q_{ij} = \frac{q}{2}(n_i n_j - \frac{\delta_{ij}}{2})}$, where $n_i$ is the director, $q$ is the magnitude of the nematic order and $\delta_{ij}$ is the Kronecker delta.
The nematic order parameter evolves according to the Beris-Edwards equation~\cite{beris_thermodynamics_1994}:
\begin{equation}
    (\partial_t + v_k \partial_k) Q_{ij} - S_{ij} = \frac{1}{\gamma} H_{ij}, \label{eq:qev}\\
\end{equation}
where $\gamma$ is the rotational diffusivity, and 
\begin{equation}
    H_{ij} = - \left( \frac{\delta \mathcal{F}}{\delta Q_{ij}}\right)^{\text{ST}} \label{eq:H}
\end{equation}
is the molecular field, describing the relaxation of the order parameter to the minimum of the free energy.
The superscript $\text{ST}$ denotes the symmetric, traceless part.
The free energy is comprised of a Landau-de Gennes term and an elastic term:
\begin{equation}
    \mathcal{F} = \int dA\, (f_{\text{LdG}} + f_{\text{el}})\,.
\end{equation}

The Landau-de Gennes term follows from the expansion of the order parameter, $Q_{ij}$:
\begin{equation}
    f_{\text{LdG}} = A(1-Q_{ij}Q_{ji})^2,
\end{equation}
where $A$ is the bulk free energy strength. 
This constant is chosen such that the nematic state is favoured.
The elastic free energy can be obtained by expanding in derivatives in $Q_{ij}$ as follows~\cite{edwards_note_1989, beris_thermodynamics_1994}:
\begin{align}
    f_{\text{el}} = &\frac{L_1}{2}\partial_kQ_{ij}\partial_kQ_{ij}\nonumber\\
    + &\frac{L_2}{2}\partial_k Q_{ik} \partial_l Q_{il}\nonumber\\
    + &\frac{L_3}{2}Q_{ij}\partial_iQ_{kl}\partial_jQ_{kl}. \label{eq:frank-3L}
\end{align}
The well-known Frank elastic energy, with splay, twist, and bend elastic constants $k_1,~k_2,~k_3$~\cite{gennes_physics_1993}:
\begin{align}
    f_{\text{Frank}} = \frac{k_1}{2} (\nabla \cdot \vec{n})^2 + \frac{k_2}{2} (\vec{n}\cdot(\nabla\times \vec{n} ))^2 + \frac{k_3}{2} (\vec{n}\times(\nabla\times \vec{n} ))^2,
\end{align}
can be recast in terms of $L_i$ from \eqref{eq:frank-3L} as~\cite{schiele_elastic_1983, beris_thermodynamics_1994, zhang_interplay_2018, kumar_catapulting_2022}:
\begin{align}
    L_1 &= \frac{2k_2-k_1+k_3}{4S^2}\nonumber\\
    L_2 &=\frac{k_1-k_2}{S^2}\nonumber\\
    L_3 &= \frac{-k_1+k_3}{2S^3}.\label{eq:Ltok}
\end{align}
When $k_1 = k_2 = k_3$, we recover the commonly used one constant approximation with elastic constant $K$~\cite{doostmohammadi_active_2018, thampi_active_2016}:
\begin{align}
    f_{el, 1} = \frac{K}{2}\partial_kQ_{ij}\partial_kQ_{ij}.
    \label{eq:fel1}
\end{align}

Finally, on the left-hand side of the order parameter evolution \eqref{eq:qev}, the co-rotation term $S_{ij}$ describes the interaction between $Q_{ij}$ and gradients in the flow $v_i$: 
\begin{align}
    S_{ij} &= (\lambda E_{ik}+\Omega_{ik})(Q_{kj}+\delta_{kj}/2) \nonumber \\
    &+(Q_{ik}+\delta_{ik}/2)(\lambda E_{kj}- \Omega_{kj})\nonumber \\
    & - 2\lambda(Q_{ij}+\delta_{ij}/2)(Q_{kl}\partial_k v_l), \label{eq:S}
\end{align}
with the rate of strain tensor ${E_{ij} = 1/2(\partial_i v_j + \partial_j v_i)}$ and vorticity tensor ${\Omega_{ij} = 1/2(\partial_i v_j - \partial_j v_i)}$.
The flow alignment parameter $\lambda$ controls the alignment of the nematic with the flow gradients, and therefore has an important role in determining the type of dynamics that arise~\cite{edwards_spontaneous_2009, thijssen_active_2020, chandragiri2019active,chandragiri2020flow}.

\begin{figure}
    \centering
    \includegraphics[width=\linewidth]{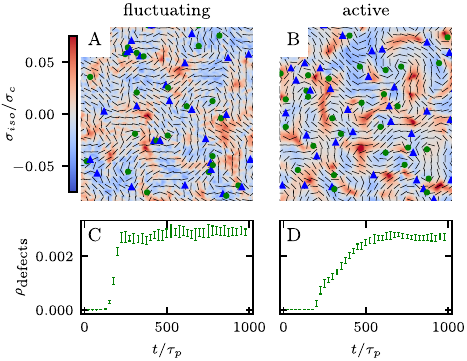}
    \caption{\textit{Defect turbulence phases for fluctuating and active nematics.}
    A and B show the defect turbulence phase for fluctuating and active nematics respectively, with directors in black, $+1/2$ defects as green circles, $-1/2$ defects as blue triangles, and the isotropic stress $\sigma_\text{iso}$ as a heat map.
    C and D show defect density for fluctuating and active nematics respectively, as a function of simulation time nondimensionalised by the passive nematic timescale $\tau_p = \gamma/A$.
    A and C are performed at $\xi_{\theta} = 0.01$. 
    B and D are performed at $\tilde \zeta = 12.5$ and both at a representative elastic constant $\tilde K=2.5$.
    The results in C and D are averages over
    $10$ replicates and error bars correspond to standard deviations.
    }
    \label{fig:snapshots}
\end{figure}

The velocity field is determined by the incompressible Navier-Stokes equations, with density $\rho$, velocity $v_i$ and generalized stress $\Pi_{ij}$:
\begin{align}
    \rho(\partial_t + v_k \partial_k)v_i = \partial_j \Pi_{ij}, \,\, \partial_k v_k = 0.\label{eq:navstok}
\end{align}
The stress, $\Pi_{ij}$, includes a pressure term, $\Pi^{p}_{ij} = -p\delta_{ij}$, viscous stresses, $\Pi^v_{ij} = -2\eta E_{ij}$ with viscosity, $\eta$, and elastic stresses $\Pi^e_{ij} = 2\lambda(Q_{ij}+\delta_{ij}/2)(Q_{lk}H_{kl}) - \lambda H_{ik}(Q_{kj}+\delta_{kj}/2)- \lambda (Q_{ik}+\delta_{ik}/2)H_{kj} + \left(f\delta_{ij} - \partial_i Q_{kl}\frac{\delta f}{\delta\partial_j Q_{lk}}\right) + Q_{ik}H_{kj} - H_{ik}Q_{kj}$, which is a function of the flow alignment parameter, $\lambda$ and the free energy density $f$.
In an active nematic, the active term $\Pi^a_{ij} = -\zeta Q_{ij}$ with activity, $\zeta$, is added.
When the activity is positive, bend perturbations in the nematic state are unstable and lead to an extensile active turbulence~\cite{doostmohammadi_active_2018}.
For a negative activity, splay perturbations are unstable and a contractile active turbulence develops. 
It is therefore expected that the response of an active nematic to varying bend and splay elastic constants will depend on the strength and sign of the active stress.
To isolate the effect of elasticity from that of active stress, we first focus on the breaking of detailed balance by nonequilibrium fluctuations, and then in section~\ref{sec:active} we study the synergistic effects of elasticity and active stress. 
Figure~\ref{fig:snapshots} shows representative snapshots of the active turbulence and defect density evolution with time, comparing systems driven out of equilibrium by fluctuations and by active stress.
\subsection{Fluctuations}
It has been shown that breaking detailed balance by fluctuations of different kinds, such as fluctuations in the order parameter field, velocity field or angle of the director, can lead to very similar defect dynamics as observed with active stresses~\cite{bonn_fluctuation-induced_2022}.

Therefore, to break detailed balance without including active stresses, we include fluctuations in the angle of the nematic.
At every step we calculate the angle $\theta_l$ of the director at lattice point $l$, and add a random angle, taken from the uniform distribution $U$, scaled by the fluctuation strength $\xi_\theta$~\cite{bonn_fluctuation-induced_2022}:
\begin{align}
    \theta_l(t+\Delta t) = \theta_l(t) + U[-\pi/2, \pi/2]\sqrt{\xi_\theta/\gamma}, \label{eq:thetanoise}
\end{align}
and then reconstruct $Q_{ij}$.
This method preserves the symmetry and tracelessness of the order parameter, changing only the effective angle.
\begin{table}[htbp]
\caption{Simulation parameters with name, symbol and value (or range) and dimension where length: $L$, mass: $M$ and time: $T$.}
\label{tab:parameters}
~\\
\resizebox{\linewidth}{!}{\begin{tabular}{@{}llll@{}}
\hline
Parameter                  & Symbol       & Value & Dimension ($2$D) \\\hline\hline
flow-alignment             & $\lambda$        & $1$  & $1$ \\
rotational viscosity     & $\gamma$     & $20$ & $M/T$\\
solvent viscosity  &        $\eta$        & $40/6$  & $M/T$\\%
density                    & $\rho$       & $40$ & $M/L^2$ \\%
bulk free energy strength         & $A$          & $1$ & $M/T^2$\\parameter
Frank elastic constant     & $K$          & $[0.01, 0.5]$  &  $ML^2/T^2$\\%
LB relaxation time         & $\tau_{LB}$       & $1$ & $T$\\%
activity                   & $\zeta$      & $[-0.05, 0.05]$ &   $M/T^2$ \\
initial noise in alignment & $n_0$        & $0.05$  & $1$ \\%
director angle fluctuation & $\xi_{\theta}$ & $0.04$ & $M/T$\\
numerical integration time & $\Delta t$   & $1$ & $T$\\%
square domain length       & $L$          & $256$   & $L$ \\\hline
\end{tabular}}
\end{table}

\subsection{Simulations setup}
We numerically solve the governing equations over a square with side length $L$ with periodic boundary conditions.
The system is initialized with a velocity field at rest and a small amount of noise in $Q_{ij}$, as in equation~\eqref{eq:thetanoise} but with $\theta(t=0)=0$.
The parameters and their values are listed in Table~\ref{tab:parameters}.
Simulations were performed over $T_\text{sim} = 3\times 10^4 \Delta t$ time steps, with a discarded warm up period of $T_\text{warmup} = 1\times 10^4 \Delta t$.
The parameters are chosen such that the system is at very low Reynolds number ($Re \sim 10^{-4}$) and the nematic coherence length scale $l_Q = \sqrt{K/A}$, is much smaller than the system size ($l_Q/L < 10^{-2}$).
Unless otherwise stated, we report simulation results in dimensionless units. To this end, as the main focus of our quantitative analyses is on the extent (area of influence) of the stresses and their intensity, we define characteristic length and characteristic stress as
$l_c = \gamma/\sqrt{(\rho A)}$, $\sigma_c = \xi_\theta A/\gamma$, respectively. Note that since our main control parameter in this study is the elastic constant the characteristic length and stress scales are defined independent of $K$.
As such, the main measurable quantities are reported as follows: the area of influence is normalized by $l_c^2$, and the stress intensity by $\sigma_c$. Similarly the control parameters are reported as follows using the same characteristic length $l_c$ and characteristic stress $\sigma_c$ scales: the dimensionless elastic constant is defined as $\tilde K = K/ (\sigma_c l_c^2)$, and the dimensionless activity as $\tilde \zeta = \zeta/ \sigma_c$. In all simulations we keep the bulk free energy coefficient $A$, constant, therefore changing elasticity $K$ affects the coherence length $\propto \sqrt{K/A}$. However, this coherence length determines the size of the defect core, i.e. the isotropic region at the core of the defect, while the extent of the stress is predominantly dependent on the average distance between the defects, which is determined by the competition of the activity (from nonequilibrium fluctuations or from active stress) and the elasticity. Furthermore, it is well-established that active defects form even if the energy constant A is zero, through the activity-induced order~\cite{thampi_intrinsic_2015}, which lends further support to the energy constant having only a minor effect on the scale of the patterns around the defect and beyond the defect core. We have further verified that doubling the bulk free energy coefficient $A$ has no significant effect on the extent and intensity of isotropic stresses around topological defects.

\section{Results}
Our focus is on the patterns of mechanical stress: the intensity and spread of stress around topological defects. To this end, we conduct simulations of active nematics, first in the presence of nonequilibrium fluctuations, and then including active stresses.
In both cases, we perform our analyses in the regime where the system is in a statistical steady state, the topological defect density is saturated and the isotropic stresses are similar (Fig.~\ref{fig:snapshots}).
In order to characterize the stress patterns we calculate the isotropic stress, i.e., the trace of the stress tensor $\sigma_\text{iso} = \frac{1}{2} \Pi_{ij}\delta_{ij}$: negative isotropic stress characterizes regions under compression and positive isotropic stress demarcates regions that are under tension. The stress tensor $\Pi_{ij}$ accounts for contributions from all components of the stress, including pressure, which ensures incompressibility condition is satisfied, viscous stresses, elastic stresses, and active stresses.
The choice of using isotropic stress is motivated by biological implications where tension and compression lead to activation-deactivation of mechanotransductive pathways~\cite{panciera2017mechanobiology, cai2021mechanoregulation} and as a result living materials respond differently to compression and tension~\cite{zulueta-coarasa_role_2022}.

\begin{figure}[htbp]
    \centering
    \includegraphics[width=\linewidth]{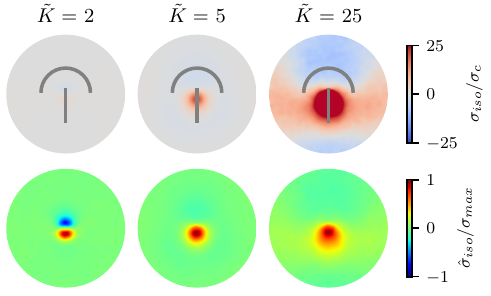}
    \caption{\textit{Isotropic stress patterns around defects for varying elasticity $K$, within one-constant approximation.}
    The upper row shows the increase in stress magnitude, growing by $2$ orders of magnitude as $K$ increases from $\tilde K=2$ to $\tilde K=25$.
    The lower row shows that while for lower elasticity the tensile and compressive regions are of similar strength and size, at higher elasticity tension dominates, changing the character of the stress pattern qualitatively. 
    The average is taken over $>1000$ defects for low, intermediate, and high elasticity values and $5$ replicates were made.
    }
    \label{fig:Kwiththeta}
\end{figure}
\subsection{Elasticity effects on stress localization: single elastic constant\label{sec:singleK}}
We begin by investigating the effect of varying the elastic constant $K$, staying within the single elastic constant approximation, on the stress patterns in fluctuating nematics.
Since elasticity penalizes deformations in the nematic director field, it is expected that the size of the ordered region in the director field around the defect increases with the increasing elasticity~\cite{kleman_topological_2006}. 
The stress generated around the defect, however, is what influences mechanics, and governs the possible role of topological defects in inducing compression-tension in the material.

At low elasticity, a highly local, dipolar stress pattern is observed with a compressive ($\sigma_\text{iso} < 0$) and a tensile ($\sigma_\text{iso} > 0$) region at the head and the tail of the comet, respectively. As such, the defect provides a local region for compression and tension, which are known to differently affect activation-deactivation of mechanosensitive pathways~\cite{panciera2017mechanobiology,zulueta-coarasa_role_2022}, with compression being linked to cell death, extrusion~\cite{saw_topological_2017}, and differentiation~\cite{guillamat2022integer}, while tension has been associated with cell division~\cite{zulueta-coarasa_role_2022}.
The compression and tension regions are symmetric around the defect core: they have the same size and have similar stress intensities (Fig. \ref{fig:Kwiththeta}).
As the elasticity is increased, the compression and tension regions respond differently and the symmetry of the isotropic stress between the head and the tail regions is lost: the tension magnitude increases faster than compression. 
The pattern then changes from a dipolar isotropic stress pattern, with equal peaks for compression and tension at the head and tail of the defect, to a pattern of a large tensile peak, but a wider and weak compressive peak the defect head.
Both tensile and compressive regions expand, but while tension localizes towards the defect core, the compression spreads around the head region of the defect.
\begin{figure}[htbp]
    \centering
    \includegraphics[width=\linewidth]{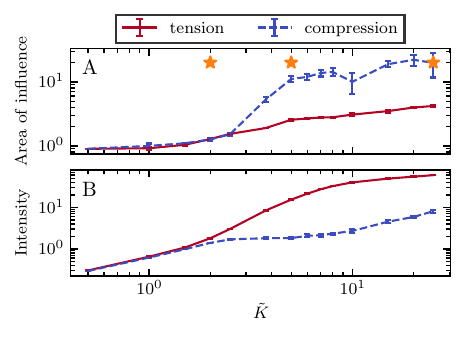}
    \caption{\textit{Intensity and extent of stresses around defects increase as elasticity is increased.}
    (A) Area of influence of tension (blue) and compression (red) regions around defect.
    At low elasticity, both peaks are of similar size, but after a threshold $K$ the compressive peak grows faster than the tensile peak, after which growth stabilizes and both peaks have a fixed difference in size.
    (B) Intensity of stress in the peak region for tension (blue) and compression (red). 
    At lower elasticity, the peaks have the same strength, however as the elasticity is increased, the tensile peak dominates. 
    Stars indicate values of $\tilde K$ used in Fig. \ref{fig:Kwiththeta}
    Simulation results are averaged over $5$ replicates. Error bars indicate standard deviation.
    }
    \label{fig:f2}
\end{figure}

To quantify the effect of elasticity on the intensity and the extent of the stresses around defects, we measure the area of influence and average magnitude of the stress, henceforth called intensity, for both the tensile and compressive regions. 
To calculate the area of influence, we first find the maximum tension and compression values and measure the size of the area over which the magnitude of the isotropic stress drops to $50\%$ of the maximum value. This gives us a measure of how the spread of both compression and tension regions change with varying the elastic constant.
Our quantitative measures clearly show that the areas of influence increase for both compression and tension as the elasticity is increased (Fig.~\ref{fig:f2}).
At a low value of the elastic constant, both the tension and compression regions have the same area of influence around the defect. 
Upon increasing the elastic constant, the size of the tensile area steadily increases, while the compression area expands to a much larger extent, to almost five-fold larger than the tensile area (Fig.~\ref{fig:f2}a).
However, measuring the intensity of the stress (Fig.~\ref{fig:f2}b) reveals that at higher elasticities the less spread, tensile area has an almost $10$-fold larger stress intensity compared to the more spread, compressive region.
Together these results show that, even within the single elastic constant approximation, changing elasticity has a significant effect on the patterns of mechanical stress localization around topological defects: while at low elasticities, both intensity and the area of influence of tensile and compressive regions are symmetric around the defect core, increasing the elastic constant breaks this symmetric localization of the stress leading to a less spread out, highly localized, tension at the defect core with a strong intensity, and a more spread out, but less intense compression at the head of the defect.
\begin{figure}[htbp]
    \centering
    \includegraphics[width=0.92\linewidth]{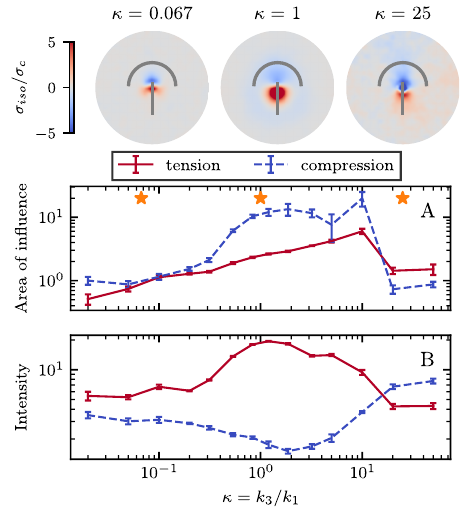}
    \caption{\textit{Isotropic stress when changing bend to splay ratio $\kappa$.}
    Top row: The normalized isotropic stress fields show a more intense tensile peak at the tail and a weaker compressive peak at the head for low $\kappa$.
    At high $\kappa$ , this effect inverts and the compression is more intense while the tension is weaker at the tail.
    (A) At low $\kappa$ the compressive and tensile peaks are similar, at intermediate $\kappa$ compression is larger, and as the ratio approaches $\kappa\approx20$ tension becomes larger.
    The stars indicate the values of $\kappa$ at which the flow fields were measured.
    (B) When splay dominates, the tensile peak is around $\approx 2$ times stronger than the compressive, with that ratio rising as $\kappa$ increases.
    However, as the ratio passes $\kappa\approx 10$, the intensity of compression becomes more dominant.
    Simulation results represent average fields over $>1000$ defects and $5$ replicates were made. 
    Error bars indicate standard deviation.
    }
    \label{fig:figk1k3imexmg}
    \label{fig:figk1k3graphs}
\end{figure}
\subsection{Elasticity effects on stress localization: distinct splay \texorpdfstring{$k_1$}{k\_1} and bend \texorpdfstring{$k_3$}{k\_3} elastic constants\label{sec:bendsplay}}
So far, we have only explored elasticity effects on stress patterns around defects, within the single elastic constant approximation, where the nematogens have the same resistance to bend and splay deformation modes. 
As mentioned earlier, this is typically not the case in passive nematics~\cite{demus2011handbook}, and it has also been shown that cytoskeletal filaments and motor protein mixtures have different bend and splay elastic constants~\cite{zhang_interplay_2018}.
Furthermore, because half-integer defects combine bend and splay deformation modes, we expect the distinction between bend and splay elastic constants to be specifically important there.
To this end, we next focus on the impact of elastic anisotropy on the patterns of isotropic stress, their intensity, and spread around topological defects.

The bend-to-splay ratio is defined as $\kappa = k_3/k_1$.
For low values of $\kappa$, when the splay elastic constant is higher, the extent of the stress pattern is small, and the has a higher intensity than the compression (Fig.~\ref{fig:figk1k3imexmg} top row).
As the bend-to-splay ratio is increased to intermediate values $\kappa \sim 1$, the tension still has higher intensity, reflecting the results from section \ref{sec:singleK}.
As the bend-to-splay ratio is increased to $\kappa\gg1$, we find that the compression is enhanced until it dominates over tension.

Similar to Sec.~\ref{sec:singleK}, the effect of changing the elastic anisotropy is best understood by quantifying the intensity and the spread of tension and compression around defects as the bend-to-splay elasticity ratio is varied (Fig. \ref{fig:figk1k3graphs}).

Three regimes are evident: low, intermediate, and high bend-to-splay ratio $\kappa$. The intermediate regime, as expected, is close to the single constant approximation, where the tension has a smaller peak with higher intensity, while the compression has a more diffuse lower peak, as described earlier.
However, the low and high $\kappa$ regimes differ substantially.
At low $\kappa$, the tensile peak is more intense but has a smaller extent, whereas at high $\kappa$ it is the compressive peak that is more intense but has a smaller area of influence (Fig.~\ref{fig:figk1k3graphs}).
This is an inversion of the effects of tension and compression, although their positioning around the defect core stays the same.

It is noteworthy that the largest stress patterns by area appear in the intermediate bend-to-splay ratio, which also has the largest intensity difference, $\approx10$-fold, whereas at low and high $\kappa$, the areas are small and the intensity differences are relatively low, $\approx2$-fold (Fig.~\ref{fig:figk1k3graphs}).

\subsection{Interplay between activity and elasticity\label{sec:active}}
In all the results presented so far, we focused on fluctuation-induced defects and purposefully neglected dipolar active stresses, which are commonly employed in models of active nematics~\cite{doostmohammadi_active_2018}. 
The purpose was to isolate the effect of elasticity from that of the active stress in inducing coherent stress patterns around topological defects.
This is especially relevant since previous theoretical and numerical works within the over-damped compressible limit, and in discrete simulations, have shown that active stresses can effectively normalize the elastic constants\cite{joshi_interplay_2019}.
Having understood the impacts of elasticity, both at the single elastic constant limit and distinct bend-splay modes, we now turn to characterizing the interplay between activity and elasticity in establishing the stress patterns, their region of influence, and intensity around topological defects. 
Motivated by experiments on eukaryotic cells~\cite{duroure2005force,duclos_topological_2017}, and actomyosin cytoskeletal filaments~\cite{clark2007myosin}, here we consider the contractile active stress, i.e. a force dipole along the nematogen's direction that is pointing inwards, and therefore contracts along the director.

\begin{figure}[htbp]
    \centering
    \includegraphics[width=\linewidth]{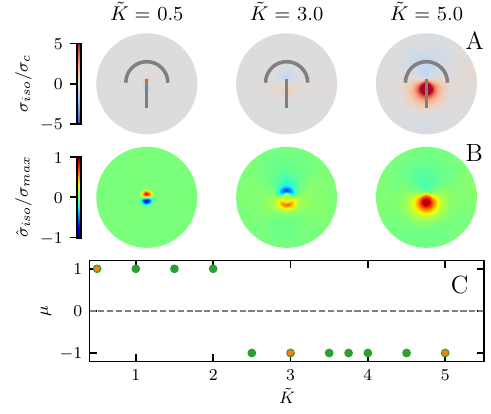}
    \caption{\textit{Isotropic stress patterns around defects for varying elasticity $K$, and a contractile active stress.}
    (A) shows the increase in stress magnitude.
    (B) shows that while for lower elasticity the compression$\rightarrow$tension dipole is pointed in the tail$\rightarrow$head direction, at higher elasticity the direction of the dipole flips to tension$\rightarrow$compression along the tail$\rightarrow$head direction. 
    (C) The sign of the stress gradient at the core of the defect changes with increasing stiffness, indicating that tension and compression have switched places. 
    Orange stars indicate points at which the stress fields are shown.
    $\tilde \zeta=-25$,
    and the averages are made over $>500$ defects and $5$ replicates.
    }
    \label{fig:posqnegz}
\end{figure}

In sections~\ref{sec:singleK} and \ref{sec:bendsplay}, we took a systematic approach of first varying only the elastic constant within the single elastic constant approximation and then analyzing the impact of elastic anisotropy. The results showed that the main trends associated with the stress intensity and extent are common within the two approaches. Thus, when we consider the interplay with activity we only focus on the single elastic constant approximation, since it is already expected that activity (depending on its sign) can favor one mode (splay or bend) over the other. Fig.~\ref{fig:posqnegz} shows the isotropic stress around a defect in a contractile active nematic, $\zeta < 0$, with nonequilibrium fluctuations, $\xi_\theta = 0.04$, within the single elastic constant approximation.
At low elasticity, the activity effect dominates and, as expected for a contractile active nematic, the compression$\rightarrow$tension dipole is aligned along the tail$\rightarrow$head direction around the comet-like $+1/2$ defect.
However, as the elasticity is increased, the compressive and tensile regions flip, resembling the stress pattern expected for an extensile active nematic, although the (contractile) active stress was kept unchanged.

A simple explanation is that while the in an active nematic the stress dipole is dominated by activity~\cite{giomi_defect_2014}, the stress localization for the fluctuation induced defect is controlled by elastic effects~\cite{bonn_fluctuation-induced_2022}.
As the elastic constant, $K$, increases, the strength of elastic effects is increased until they overcome the active effects, and the stress dipole flips direction from compression$\rightarrow$tension dipole to tension$\rightarrow$compression along the tail$\rightarrow$head direction around the comet-like $+1/2$ defect.
This switch in the tension-compression regions by only changing the elastic constant presents an interesting, simple, way for switching the dynamic behavior in active matter by changing a passive material property. It is also noteworthy that this flipping of the stress also corresponds to the switch between extensile and contractile behavior of topological defects in active nematics~\cite{doostmohammadi_active_2018}. Experimental evidence of such a switch is presented in epithelial layers upon weakening of cell-cell adhesion~\cite{balasubramaniam_investigating_2021}, and while various active mechanisms have been proposed for the flipping of extensile-contractile defect behavior~\cite{killeen_polar_2022,bonn_fluctuation-induced_2022}, these results show that such a flipping can simply occur upon changing of the oreintational elasticity of the nematogens, a passive material property. 

To characterize the flipping of tension-compression regions clearly, we measure the stress gradient along the tail$\rightarrow$head direction around the defect core $\mathcal{S} = \max_{\text{tail} \rightarrow \text{head}}\langle \frac{\partial \sigma^{\text{iso.}}}{ \partial_{\text{tail} \rightarrow \text{head}}}\rangle_\bot$, where the average is taken perpendicular to the defect axis and the maximum gradient is picked along the defect axis, and define
\begin{eqnarray}
    \mu = \frac{\mathcal{S}}{\text{max}(|\mathcal{S}|)},
\end{eqnarray}
such that $\mu < 0$, characterizes tension localized at the tail and compression at the head of the defect, while $\mu > 0$ corresponds to the reverse localization.
We see the value of $\mu$ switch to negative at $\tilde K \approx2.2$ (Fig. \ref{fig:posqnegz} C) although this value is dependent on fluctuation and active stress strength.
Having shown the effect of changing elasticity in a nematic with active fluctuations and stresses, we then construct a phase diagram of the system behavior to summarize the effect of elasticity and active stress.

\begin{figure}[htbp]
    \centering
    \includegraphics{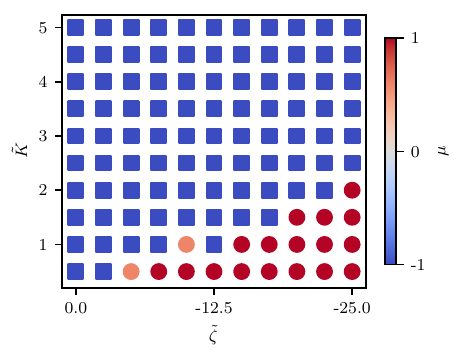}
    \caption{\textit{Phase diagram of the stress gradient around defects in the orientational elasticity-activity phase space.} Varying either elasticity or activity can lead to flipping of the direction from tension$\rightarrow$compression (blue square) dipole to compression$\rightarrow$tension along the tail$\rightarrow$head direction (red circle) around the comet-like $+1/2$ defect.
    Each point is averaged over $5$ realizations.
    }
    \label{fig:sigmaphasediag}
\end{figure}

We find that the phase boundary between tension at the head and tension at the tail is determined both by the elasticity and the activity as $\sim K/\zeta$ (Fig.~\ref{fig:sigmaphasediag}).
The phase can therefore be changed by varying either of these parameters, both of which are experimentally accessible.

\subsection{Elasticity effects on compression-tension patterns around $-1/2$ defects\label{sec:neg}}

\begin{figure}[htbp!]
    \centering
    \includegraphics[width=\linewidth]{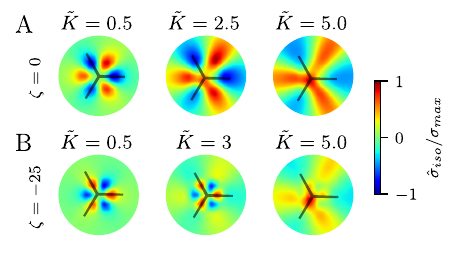}
    \caption{\textit{Isotropic stress patterns around $-1/2$ defects for varying single elastic constant, $K$.}
    (A) In the presence of nonequilibrium fluctuations, for lower values of elasticity, the bend regions around $-1/2$ are under tension.
    As the elasticity is increased, the area of tension in bend regions increases.
    (B) In the active contractile system, at low $K$ the bend regions are compressive, whereas at higher $K$ the bend regions are tensile. The averages are made over 5 simulations.
    }
    \label{fig:negdef}
\end{figure}

Finally, we explore the effect of varying the single elastic constant, $K$, on the trefoil-like, $-1/2$ defects.
Since the single and multiple elastic constants approaches give qualitatively similar results, here we focus on the single elastic constant case. 
Fig. \ref{fig:negdef}A shows the isotropic stress around a $-1/2$ defect in a passive system ($\zeta=0$) with nonequilibrium fluctuations. 
When the elastic constant is low, the bend regions are under tension, whereas splay regions experience compression.
As the elastic constant increases the area of tension in bend regions also increases.
For the contractile system ($\zeta < 0$) with nonequilibrium fluctuations the bend regions are compressive for low elasticities (Fig. \ref{fig:negdef}B).
This changes when the elastic constant is increased, where the bend regions are now under tension.
This stress inversion is reminiscent of our earlier finding of the inversion of the isotropic stress dipole pattern around $+1/2$ defects upon increasing the elastic constant for a contractile and fluctuating system.

\section{Discussion\label{sec:disc}}

Topological defects in the alignment of living entities have been found to play a functional role in a range of biological behaviors by providing local environments for enhanced mechanical stresses~\cite{saw_topological_2017,doostmohammadi_physics_2021,shankar2022topological}.
In this work, we have taken a systematic approach of exploring intensity and area of influence of mechanical stresses around topological defects.
To disentangle the effect of self-generated active stresses from passive material properties we separately investigated stress patterns in out-of-equilibrium fluctuating nematics and in active nematics.
Our results show that in both cases, the orientational elasticity of nematogens plays a significant role in determining the stress localization patterns around defects.
Due to the generic presence of fluctuations in active nematic systems, from subcellular fluctuations in focal adhesion~\cite{plotnikov_force_2012} or stress fibers~\cite{guolla_force_2012} to cell level shape fluctuations in MDCK layers~\cite{zehnder_cell_2015} or \textit{Drosophila}~\cite{olenik2023fluctuations}, we expect that understanding the interplay between activity and fluctuations will play an important role in understanding cell behavior in the future.

First, in a fluctuating system, we showed that increasing elasticity breaks the symmetry between compressive and tensile regions at the head and the tail of the defect, respectively. 
While the area of influence of the stress and the stress intensity increase for both compression and tension regions, we find that such enhancement occurs disproportionately: upon increasing elasticity, the region of tension localizes close to the defect core and the tension intensity becomes about an order of magnitude stronger than the compression.
At the same time, the compressive region spreads over a relatively larger region around the defect compared to tension.
In MDCK cells, knockdown of the mechanosensor $\alpha$-catenin, and therefore weakening cell-cell contacts, was shown to reduce the overall extent of the isotropic stress patterns around defects, which was interpreted as lowering the elastic constant~\cite{saw_topological_2017}.

Upon relaxing the one-constant approximation and allowing for different bend and splay elasticities, we find that for low and intermediate bend to splay ratios, the stress patterns around defects are characterized by strong, highly local tension and weak, more spread out compression.
However, this behavior switches for high bend to splay ratios where local strong compression and wide-spread weak tension regions emerge.

Finally, combining the effects of activity and elasticity, we present the possibility of the modulation of stress intensity, area of the influence of stress, and also flipping of the compression-tension regions. 
This inversion of compressive and tensile regions was observed in both +1/2 and -1/2 defects.
Taken together, the results show that in a nonequilibrium fluctuating nematic system, we saw that changing the elasticity can change the dipole stress pattern to increase the intensity of tension and the area of influence of compression.
In an active and nonequilibrium fluctuating nematic system, we found that changing the elasticity can even lead to the flipping of the localization of compression-tension regions around the defects.

The results presented herein can be useful in understanding how stress localization is controlled by the self-organization of the active entities. Understanding the strength and extent of the localization of stresses that collections of elongated active entities are subject to, and how they depend on material properties of the entities themselves, is essential in studying the mechanically-induced biochemical changes in living biological matter.
For example, the compressive region of the isotropic stress pattern of a topological defect has been found to lead to cell death and extrusion in mammalian epithelia~\cite{saw_topological_2017} and tension in a tissue is associated with cell division~\cite{zulueta-coarasa_role_2022} or gap opening on soft substrates~\cite{sonam_2023_mechanical}.
Indeed, the existing experiments that have measured the compression-tension patterns around the defects, have shown that molecular perturbation of cell-cell contacts, which inevitably alters the mechanical properties of the tissues, can affect the intensity and extent of isotropic stress localization around defects~\cite{saw_topological_2017}, and could even reverse the direction of the motion of defects~\cite{balasubramaniam_investigating_2021}.

Experimentally, varying the bend to splay constant is possible, as simply varying the length of a flexible nematic such as actin has been shown to change the bend to splay ratio~\cite{joshi_interplay_2019, zhang_interplay_2018}.
Furthermore, adding carbon nanotubes to a molecular nematic was shown to increase the bend to splay ratio by up to $6$ fold~\cite{turlapati2017elastic}, while in a similar fashion, microtubules were added to an active nematic consisting of actin and myosin to increase the bend constant $>2$-fold~\cite{zhang_interplay_2018}.
We predict that the change from intense tension to intense compression when varying the bend to splay ratio should affect cellular systems, as tension and compression are linked to differing biological behaviors~\cite{zulueta-coarasa_role_2022}.

We expect our results to help design new experiments aiming at controlling stress localization, and potentially mechanotransduction, by adjusting the elastic properties of the cell layers.

\section*{Conflicts of interest}
There are no conflicts to declare.

\section*{Acknowledgements}
A. D. acknowledges funding from the Novo Nordisk Foundation (grant No. NNF18SA0035142 and NERD grant No. NNF21OC0068687), Villum Fonden (Grant no. 29476), and the European Union (ERC, PhysCoMeT, 101041418). Views and opinions expressed are however those of the authors only and do not necessarily reflect those of the European Union or the European Research Council. Neither the European Union nor the granting authority can be held responsible for them. A.A. acknowledges support from the European Union’s Horizon Europe research and innovation program under the Marie Sklodowska-Curie grant agreement No. 101063870 (TopCellComm).



\balance


\bibliography{activematter} 

\providecommand*{\mcitethebibliography}{\thebibliography}
\csname @ifundefined\endcsname{endmcitethebibliography}
{\let\endmcitethebibliography\endthebibliography}{}
\begin{mcitethebibliography}{56}
\providecommand*{\natexlab}[1]{#1}
\providecommand*{\mciteSetBstSublistMode}[1]{}
\providecommand*{\mciteSetBstMaxWidthForm}[2]{}
\providecommand*{\mciteBstWouldAddEndPuncttrue}
  {\def\EndOfBibitem{\unskip.}}
\providecommand*{\mciteBstWouldAddEndPunctfalse}
  {\let\EndOfBibitem\relax}
\providecommand*{\mciteSetBstMidEndSepPunct}[3]{}
\providecommand*{\mciteSetBstSublistLabelBeginEnd}[3]{}
\providecommand*{\EndOfBibitem}{}
\mciteSetBstSublistMode{f}
\mciteSetBstMaxWidthForm{subitem}
{(\emph{\alph{mcitesubitemcount}})}
\mciteSetBstSublistLabelBeginEnd{\mcitemaxwidthsubitemform\space}
{\relax}{\relax}

\bibitem[Hayward \emph{et~al.}(2021)Hayward, Muncie, and Weaver]{hayward2021tissue}
M.-K. Hayward, J.~M. Muncie and V.~M. Weaver, \emph{Developmental cell}, 2021, \textbf{56}, 1833--1847\relax
\mciteBstWouldAddEndPuncttrue
\mciteSetBstMidEndSepPunct{\mcitedefaultmidpunct}
{\mcitedefaultendpunct}{\mcitedefaultseppunct}\relax
\EndOfBibitem
\bibitem[Vining and Mooney(2017)]{vining2017mechanical}
K.~H. Vining and D.~J. Mooney, \emph{Nature reviews Molecular cell biology}, 2017, \textbf{18}, 728--742\relax
\mciteBstWouldAddEndPuncttrue
\mciteSetBstMidEndSepPunct{\mcitedefaultmidpunct}
{\mcitedefaultendpunct}{\mcitedefaultseppunct}\relax
\EndOfBibitem
\bibitem[Ladoux and Mège(2017)]{ladoux_mechanobiology_2017}
B.~Ladoux and R.-M. Mège, \emph{Nature Reviews Molecular Cell Biology}, 2017, \textbf{18}, 743--757\relax
\mciteBstWouldAddEndPuncttrue
\mciteSetBstMidEndSepPunct{\mcitedefaultmidpunct}
{\mcitedefaultendpunct}{\mcitedefaultseppunct}\relax
\EndOfBibitem
\bibitem[Persat \emph{et~al.}(2015)Persat, Nadell, Kim, Ingremeau, Siryaporn, Drescher, Wingreen, Bassler, Gitai, and Stone]{persat2015mechanical}
A.~Persat, C.~D. Nadell, M.~K. Kim, F.~Ingremeau, A.~Siryaporn, K.~Drescher, N.~S. Wingreen, B.~L. Bassler, Z.~Gitai and H.~A. Stone, \emph{Cell}, 2015, \textbf{161}, 988--997\relax
\mciteBstWouldAddEndPuncttrue
\mciteSetBstMidEndSepPunct{\mcitedefaultmidpunct}
{\mcitedefaultendpunct}{\mcitedefaultseppunct}\relax
\EndOfBibitem
\bibitem[Wang(2017)]{wang2017review}
N.~Wang, \emph{Journal of physics D: Applied physics}, 2017, \textbf{50}, 233002\relax
\mciteBstWouldAddEndPuncttrue
\mciteSetBstMidEndSepPunct{\mcitedefaultmidpunct}
{\mcitedefaultendpunct}{\mcitedefaultseppunct}\relax
\EndOfBibitem
\bibitem[Dufr{\^e}ne and Persat(2020)]{dufrene2020mechanomicrobiology}
Y.~F. Dufr{\^e}ne and A.~Persat, \emph{Nature Reviews Microbiology}, 2020, \textbf{18}, 227--240\relax
\mciteBstWouldAddEndPuncttrue
\mciteSetBstMidEndSepPunct{\mcitedefaultmidpunct}
{\mcitedefaultendpunct}{\mcitedefaultseppunct}\relax
\EndOfBibitem
\bibitem[Pocaterra \emph{et~al.}(2020)Pocaterra, Romani, and Dupont]{pocaterra_yaptaz_2020}
A.~Pocaterra, P.~Romani and S.~Dupont, \emph{Journal of Cell Science}, 2020, \textbf{133}, jcs230425\relax
\mciteBstWouldAddEndPuncttrue
\mciteSetBstMidEndSepPunct{\mcitedefaultmidpunct}
{\mcitedefaultendpunct}{\mcitedefaultseppunct}\relax
\EndOfBibitem
\bibitem[Panciera \emph{et~al.}(2017)Panciera, Azzolin, Cordenonsi, and Piccolo]{panciera2017mechanobiology}
T.~Panciera, L.~Azzolin, M.~Cordenonsi and S.~Piccolo, \emph{Nature reviews Molecular cell biology}, 2017, \textbf{18}, 758--770\relax
\mciteBstWouldAddEndPuncttrue
\mciteSetBstMidEndSepPunct{\mcitedefaultmidpunct}
{\mcitedefaultendpunct}{\mcitedefaultseppunct}\relax
\EndOfBibitem
\bibitem[Mathieu and Manneville(2019)]{mathieu2019intracellular}
S.~Mathieu and J.-B. Manneville, \emph{Current opinion in cell biology}, 2019, \textbf{56}, 34--44\relax
\mciteBstWouldAddEndPuncttrue
\mciteSetBstMidEndSepPunct{\mcitedefaultmidpunct}
{\mcitedefaultendpunct}{\mcitedefaultseppunct}\relax
\EndOfBibitem
\bibitem[Boocock \emph{et~al.}(2023)Boocock, Hirashima, and Hannezo]{boocock2023interplay}
D.~Boocock, T.~Hirashima and E.~Hannezo, \emph{bioRxiv}, 2023,  2023--03\relax
\mciteBstWouldAddEndPuncttrue
\mciteSetBstMidEndSepPunct{\mcitedefaultmidpunct}
{\mcitedefaultendpunct}{\mcitedefaultseppunct}\relax
\EndOfBibitem
\bibitem[Brugu{\'e}s \emph{et~al.}(2014)Brugu{\'e}s, Anon, Conte, Veldhuis, Gupta, Colombelli, Mu{\~n}oz, Brodland, Ladoux, and Trepat]{brugues2014forces}
A.~Brugu{\'e}s, E.~Anon, V.~Conte, J.~H. Veldhuis, M.~Gupta, J.~Colombelli, J.~J. Mu{\~n}oz, G.~W. Brodland, B.~Ladoux and X.~Trepat, \emph{Nature physics}, 2014, \textbf{10}, 683--690\relax
\mciteBstWouldAddEndPuncttrue
\mciteSetBstMidEndSepPunct{\mcitedefaultmidpunct}
{\mcitedefaultendpunct}{\mcitedefaultseppunct}\relax
\EndOfBibitem
\bibitem[Monfared \emph{et~al.}(2023)Monfared, Ravichandran, Andrade, and Doostmohammadi]{monfared2023mechanical}
S.~Monfared, G.~Ravichandran, J.~Andrade and A.~Doostmohammadi, \emph{Elife}, 2023, \textbf{12}, e82435\relax
\mciteBstWouldAddEndPuncttrue
\mciteSetBstMidEndSepPunct{\mcitedefaultmidpunct}
{\mcitedefaultendpunct}{\mcitedefaultseppunct}\relax
\EndOfBibitem
\bibitem[Saw \emph{et~al.}(2017)Saw, Doostmohammadi, Nier, Kocgozlu, Thampi, Toyama, Marcq, Lim, Yeomans, and Ladoux]{saw_topological_2017}
T.~B. Saw, A.~Doostmohammadi, V.~Nier, L.~Kocgozlu, S.~Thampi, Y.~Toyama, P.~Marcq, C.~T. Lim, J.~M. Yeomans and B.~Ladoux, \emph{Nature}, 2017, \textbf{544}, 212--216\relax
\mciteBstWouldAddEndPuncttrue
\mciteSetBstMidEndSepPunct{\mcitedefaultmidpunct}
{\mcitedefaultendpunct}{\mcitedefaultseppunct}\relax
\EndOfBibitem
\bibitem[Doostmohammadi and Ladoux(2021)]{doostmohammadi_physics_2021}
A.~Doostmohammadi and B.~Ladoux, \emph{Trends in Cell Biology}, 2021\relax
\mciteBstWouldAddEndPuncttrue
\mciteSetBstMidEndSepPunct{\mcitedefaultmidpunct}
{\mcitedefaultendpunct}{\mcitedefaultseppunct}\relax
\EndOfBibitem
\bibitem[Kawaguchi \emph{et~al.}(2017)Kawaguchi, Kageyama, and Sano]{kawaguchi_topological_2017}
K.~Kawaguchi, R.~Kageyama and M.~Sano, \emph{Nature}, 2017, \textbf{545}, 327--331\relax
\mciteBstWouldAddEndPuncttrue
\mciteSetBstMidEndSepPunct{\mcitedefaultmidpunct}
{\mcitedefaultendpunct}{\mcitedefaultseppunct}\relax
\EndOfBibitem
\bibitem[Meacock \emph{et~al.}(2021)Meacock, Doostmohammadi, Foster, Yeomans, and Durham]{meacock_bacteria_2021}
O.~J. Meacock, A.~Doostmohammadi, K.~R. Foster, J.~M. Yeomans and W.~M. Durham, \emph{Nature Physics}, 2021, \textbf{17}, 205--210\relax
\mciteBstWouldAddEndPuncttrue
\mciteSetBstMidEndSepPunct{\mcitedefaultmidpunct}
{\mcitedefaultendpunct}{\mcitedefaultseppunct}\relax
\EndOfBibitem
\bibitem[Maroudas-Sacks \emph{et~al.}(2021)Maroudas-Sacks, Garion, Shani-Zerbib, Livshits, Braun, and Keren]{maroudas-sacks_topological_2021}
Y.~Maroudas-Sacks, L.~Garion, L.~Shani-Zerbib, A.~Livshits, E.~Braun and K.~Keren, \emph{Nature Physics}, 2021, \textbf{17}, 251--259\relax
\mciteBstWouldAddEndPuncttrue
\mciteSetBstMidEndSepPunct{\mcitedefaultmidpunct}
{\mcitedefaultendpunct}{\mcitedefaultseppunct}\relax
\EndOfBibitem
\bibitem[Guillamat \emph{et~al.}(2022)Guillamat, Blanch-Mercader, Pernollet, Kruse, and Roux]{guillamat2022integer}
P.~Guillamat, C.~Blanch-Mercader, G.~Pernollet, K.~Kruse and A.~Roux, \emph{Nature materials}, 2022, \textbf{21}, 588--597\relax
\mciteBstWouldAddEndPuncttrue
\mciteSetBstMidEndSepPunct{\mcitedefaultmidpunct}
{\mcitedefaultendpunct}{\mcitedefaultseppunct}\relax
\EndOfBibitem
\bibitem[Gennes and Prost(1993)]{gennes_physics_1993}
P.~G.~d. Gennes and J.~Prost, \emph{The physics of liquid crystals}, Clarendon Press ; Oxford University Press, Oxford : New York, 2nd edn, 1993\relax
\mciteBstWouldAddEndPuncttrue
\mciteSetBstMidEndSepPunct{\mcitedefaultmidpunct}
{\mcitedefaultendpunct}{\mcitedefaultseppunct}\relax
\EndOfBibitem
\bibitem[Doostmohammadi \emph{et~al.}(2018)Doostmohammadi, Ignés-Mullol, Yeomans, and Sagués]{doostmohammadi_active_2018}
A.~Doostmohammadi, J.~Ignés-Mullol, J.~M. Yeomans and F.~Sagués, \emph{Nature Communications}, 2018, \textbf{9}, 3246\relax
\mciteBstWouldAddEndPuncttrue
\mciteSetBstMidEndSepPunct{\mcitedefaultmidpunct}
{\mcitedefaultendpunct}{\mcitedefaultseppunct}\relax
\EndOfBibitem
\bibitem[Sonam \emph{et~al.}(2023)Sonam, Balasubramaniam, Lin, Ivan, Pi-Jaum{\`a}, Jebane, Karnat, Toyama, Marcq, Prost,\emph{et~al.}]{sonam_2023_mechanical}
S.~Sonam, L.~Balasubramaniam, S.-Z. Lin, Y.~M.~Y. Ivan, I.~Pi-Jaum{\`a}, C.~Jebane, M.~Karnat, Y.~Toyama, P.~Marcq, J.~Prost \emph{et~al.}, \emph{Nature Physics}, 2023, \textbf{19}, 132--141\relax
\mciteBstWouldAddEndPuncttrue
\mciteSetBstMidEndSepPunct{\mcitedefaultmidpunct}
{\mcitedefaultendpunct}{\mcitedefaultseppunct}\relax
\EndOfBibitem
\bibitem[Zulueta-Coarasa and Rosenblatt(2022)]{zulueta-coarasa_role_2022}
T.~Zulueta-Coarasa and J.~Rosenblatt, \emph{Current Opinion in Genetics \& Development}, 2022, \textbf{72}, 1--7\relax
\mciteBstWouldAddEndPuncttrue
\mciteSetBstMidEndSepPunct{\mcitedefaultmidpunct}
{\mcitedefaultendpunct}{\mcitedefaultseppunct}\relax
\EndOfBibitem
\bibitem[Tóth \emph{et~al.}(2002)Tóth, Denniston, and Yeomans]{toth_hydrodynamics_2002}
G.~Tóth, C.~Denniston and J.~M. Yeomans, \emph{Physical Review Letters}, 2002, \textbf{88}, 105504\relax
\mciteBstWouldAddEndPuncttrue
\mciteSetBstMidEndSepPunct{\mcitedefaultmidpunct}
{\mcitedefaultendpunct}{\mcitedefaultseppunct}\relax
\EndOfBibitem
\bibitem[Khoromskaia and Alexander(2017)]{khoromskaia2017vortex}
D.~Khoromskaia and G.~P. Alexander, \emph{New Journal of Physics}, 2017, \textbf{19}, 103043\relax
\mciteBstWouldAddEndPuncttrue
\mciteSetBstMidEndSepPunct{\mcitedefaultmidpunct}
{\mcitedefaultendpunct}{\mcitedefaultseppunct}\relax
\EndOfBibitem
\bibitem[Bonn \emph{et~al.}(2022)Bonn, Ardaševa, Mueller, Shendruk, and Doostmohammadi]{bonn_fluctuation-induced_2022}
L.~Bonn, A.~Ardaševa, R.~Mueller, T.~N. Shendruk and A.~Doostmohammadi, \emph{Physical Review E}, 2022, \textbf{106}, 044706\relax
\mciteBstWouldAddEndPuncttrue
\mciteSetBstMidEndSepPunct{\mcitedefaultmidpunct}
{\mcitedefaultendpunct}{\mcitedefaultseppunct}\relax
\EndOfBibitem
\bibitem[Zhang \emph{et~al.}(2018)Zhang, Kumar, Ross, Gardel, and de~Pablo]{zhang_interplay_2018}
R.~Zhang, N.~Kumar, J.~L. Ross, M.~L. Gardel and J.~J. de~Pablo, \emph{Proceedings of the National Academy of Sciences}, 2018, \textbf{115}, E124--E133\relax
\mciteBstWouldAddEndPuncttrue
\mciteSetBstMidEndSepPunct{\mcitedefaultmidpunct}
{\mcitedefaultendpunct}{\mcitedefaultseppunct}\relax
\EndOfBibitem
\bibitem[Kumar \emph{et~al.}(2018)Kumar, Zhang, de~Pablo, and Gardel]{kumar_tunable_2018}
N.~Kumar, R.~Zhang, J.~J. de~Pablo and M.~L. Gardel, \emph{Science Advances}, 2018, \textbf{4}, eaat7779\relax
\mciteBstWouldAddEndPuncttrue
\mciteSetBstMidEndSepPunct{\mcitedefaultmidpunct}
{\mcitedefaultendpunct}{\mcitedefaultseppunct}\relax
\EndOfBibitem
\bibitem[Joshi \emph{et~al.}(2019)Joshi, Putzig, Baskaran, and F. Hagan]{joshi_interplay_2019}
A.~Joshi, E.~Putzig, A.~Baskaran and M.~F. Hagan, \emph{Soft Matter}, 2019, \textbf{15}, 94--101\relax
\mciteBstWouldAddEndPuncttrue
\mciteSetBstMidEndSepPunct{\mcitedefaultmidpunct}
{\mcitedefaultendpunct}{\mcitedefaultseppunct}\relax
\EndOfBibitem
\bibitem[Thampi \emph{et~al.}(2014)Thampi, Golestanian, and Yeomans]{thampi_instabilities_2014}
S.~P. Thampi, R.~Golestanian and J.~M. Yeomans, \emph{EPL (Europhysics Letters)}, 2014, \textbf{105}, 18001\relax
\mciteBstWouldAddEndPuncttrue
\mciteSetBstMidEndSepPunct{\mcitedefaultmidpunct}
{\mcitedefaultendpunct}{\mcitedefaultseppunct}\relax
\EndOfBibitem
\bibitem[Khoromskaia and Salbreux(2023)]{khoromskaia_active_2023}
D.~Khoromskaia and G.~Salbreux, \emph{eLife}, 2023, \textbf{12}, e75878\relax
\mciteBstWouldAddEndPuncttrue
\mciteSetBstMidEndSepPunct{\mcitedefaultmidpunct}
{\mcitedefaultendpunct}{\mcitedefaultseppunct}\relax
\EndOfBibitem
\bibitem[Marenduzzo \emph{et~al.}(2007)Marenduzzo, Orlandini, Cates, and Yeomans]{Marenduzzo_hlb_2007}
D.~Marenduzzo, E.~Orlandini, M.~E. Cates and J.~M. Yeomans, \emph{Physical Review E}, 2007, \textbf{76}, 031921\relax
\mciteBstWouldAddEndPuncttrue
\mciteSetBstMidEndSepPunct{\mcitedefaultmidpunct}
{\mcitedefaultendpunct}{\mcitedefaultseppunct}\relax
\EndOfBibitem
\bibitem[Beris and Edwards(1994)]{beris_thermodynamics_1994}
A.~N. Beris and B.~J. Edwards, \emph{Thermodynamics of flowing systems: with internal microstructure}, Oxford University Press, New York, 1994\relax
\mciteBstWouldAddEndPuncttrue
\mciteSetBstMidEndSepPunct{\mcitedefaultmidpunct}
{\mcitedefaultendpunct}{\mcitedefaultseppunct}\relax
\EndOfBibitem
\bibitem[Edwards and Beris(1989)]{edwards_note_1989}
B.~J. Edwards and A.~N. Beris, \emph{Journal of Rheology}, 1989, \textbf{33}, 1189--1193\relax
\mciteBstWouldAddEndPuncttrue
\mciteSetBstMidEndSepPunct{\mcitedefaultmidpunct}
{\mcitedefaultendpunct}{\mcitedefaultseppunct}\relax
\EndOfBibitem
\bibitem[Schiele and Trimper(1983)]{schiele_elastic_1983}
K.~Schiele and S.~Trimper, \emph{physica status solidi (b)}, 1983, \textbf{118}, 267--274\relax
\mciteBstWouldAddEndPuncttrue
\mciteSetBstMidEndSepPunct{\mcitedefaultmidpunct}
{\mcitedefaultendpunct}{\mcitedefaultseppunct}\relax
\EndOfBibitem
\bibitem[Kumar \emph{et~al.}(2022)Kumar, Zhang, Redford, Pablo, and Gardel]{kumar_catapulting_2022}
N.~Kumar, R.~Zhang, S.~A. Redford, J.~J.~d. Pablo and M.~L. Gardel, \emph{Soft Matter}, 2022, \textbf{18}, 5271--5281\relax
\mciteBstWouldAddEndPuncttrue
\mciteSetBstMidEndSepPunct{\mcitedefaultmidpunct}
{\mcitedefaultendpunct}{\mcitedefaultseppunct}\relax
\EndOfBibitem
\bibitem[Thampi and Yeomans(2016)]{thampi_active_2016}
S.~Thampi and J.~Yeomans, \emph{The European Physical Journal Special Topics}, 2016, \textbf{225}, 651--662\relax
\mciteBstWouldAddEndPuncttrue
\mciteSetBstMidEndSepPunct{\mcitedefaultmidpunct}
{\mcitedefaultendpunct}{\mcitedefaultseppunct}\relax
\EndOfBibitem
\bibitem[Edwards and Yeomans(2009)]{edwards_spontaneous_2009}
S.~A. Edwards and J.~M. Yeomans, \emph{EPL (Europhysics Letters)}, 2009, \textbf{85}, 18008\relax
\mciteBstWouldAddEndPuncttrue
\mciteSetBstMidEndSepPunct{\mcitedefaultmidpunct}
{\mcitedefaultendpunct}{\mcitedefaultseppunct}\relax
\EndOfBibitem
\bibitem[Thijssen \emph{et~al.}(2020)Thijssen, Metselaar, M. Yeomans, and Doostmohammadi]{thijssen_active_2020}
K.~Thijssen, L.~Metselaar, J.~M. Yeomans and A.~Doostmohammadi, \emph{Soft Matter}, 2020, \textbf{16}, 2065--2074\relax
\mciteBstWouldAddEndPuncttrue
\mciteSetBstMidEndSepPunct{\mcitedefaultmidpunct}
{\mcitedefaultendpunct}{\mcitedefaultseppunct}\relax
\EndOfBibitem
\bibitem[Chandragiri \emph{et~al.}(2019)Chandragiri, Doostmohammadi, Yeomans, and Thampi]{chandragiri2019active}
S.~Chandragiri, A.~Doostmohammadi, J.~M. Yeomans and S.~P. Thampi, \emph{Soft matter}, 2019, \textbf{15}, 1597--1604\relax
\mciteBstWouldAddEndPuncttrue
\mciteSetBstMidEndSepPunct{\mcitedefaultmidpunct}
{\mcitedefaultendpunct}{\mcitedefaultseppunct}\relax
\EndOfBibitem
\bibitem[Chandragiri \emph{et~al.}(2020)Chandragiri, Doostmohammadi, Yeomans, and Thampi]{chandragiri2020flow}
S.~Chandragiri, A.~Doostmohammadi, J.~M. Yeomans and S.~P. Thampi, \emph{Physical Review Letters}, 2020, \textbf{125}, 148002\relax
\mciteBstWouldAddEndPuncttrue
\mciteSetBstMidEndSepPunct{\mcitedefaultmidpunct}
{\mcitedefaultendpunct}{\mcitedefaultseppunct}\relax
\EndOfBibitem
\bibitem[Thampi \emph{et~al.}(2015)Thampi, Doostmohammadi, Golestanian, and Yeomans]{thampi_intrinsic_2015}
S.~P. Thampi, A.~Doostmohammadi, R.~Golestanian and J.~M. Yeomans, \emph{EPL (Europhysics Letters)}, 2015, \textbf{112}, 28004\relax
\mciteBstWouldAddEndPuncttrue
\mciteSetBstMidEndSepPunct{\mcitedefaultmidpunct}
{\mcitedefaultendpunct}{\mcitedefaultseppunct}\relax
\EndOfBibitem
\bibitem[Cai \emph{et~al.}(2021)Cai, Wang, and Meng]{cai2021mechanoregulation}
X.~Cai, K.-C. Wang and Z.~Meng, \emph{Frontiers in cell and developmental biology}, 2021, \textbf{9}, 673599\relax
\mciteBstWouldAddEndPuncttrue
\mciteSetBstMidEndSepPunct{\mcitedefaultmidpunct}
{\mcitedefaultendpunct}{\mcitedefaultseppunct}\relax
\EndOfBibitem
\bibitem[Kleman and Lavrentovich(2006)]{kleman_topological_2006}
M.~Kleman and O.~D. Lavrentovich, \emph{Philosophical Magazine}, 2006, \textbf{86}, 4117--4137\relax
\mciteBstWouldAddEndPuncttrue
\mciteSetBstMidEndSepPunct{\mcitedefaultmidpunct}
{\mcitedefaultendpunct}{\mcitedefaultseppunct}\relax
\EndOfBibitem
\bibitem[Demus \emph{et~al.}(2011)Demus, Goodby, Gray, Spiess, and Vill]{demus2011handbook}
D.~Demus, J.~W. Goodby, G.~W. Gray, H.~W. Spiess and V.~Vill, \emph{Handbook of liquid crystals, volume 2A: low molecular weight liquid crystals I: calamitic liquid crystals}, John Wiley \& Sons, 2011\relax
\mciteBstWouldAddEndPuncttrue
\mciteSetBstMidEndSepPunct{\mcitedefaultmidpunct}
{\mcitedefaultendpunct}{\mcitedefaultseppunct}\relax
\EndOfBibitem
\bibitem[Du~Roure \emph{et~al.}(2005)Du~Roure, Saez, Buguin, Austin, Chavrier, Siberzan, and Ladoux]{duroure2005force}
O.~Du~Roure, A.~Saez, A.~Buguin, R.~H. Austin, P.~Chavrier, P.~Siberzan and B.~Ladoux, \emph{Proceedings of the National Academy of Sciences}, 2005, \textbf{102}, 2390--2395\relax
\mciteBstWouldAddEndPuncttrue
\mciteSetBstMidEndSepPunct{\mcitedefaultmidpunct}
{\mcitedefaultendpunct}{\mcitedefaultseppunct}\relax
\EndOfBibitem
\bibitem[Duclos \emph{et~al.}(2017)Duclos, Erlenkämper, Joanny, and Silberzan]{duclos_topological_2017}
G.~Duclos, C.~Erlenkämper, J.-F. Joanny and P.~Silberzan, \emph{Nature Physics}, 2017, \textbf{13}, 58--62\relax
\mciteBstWouldAddEndPuncttrue
\mciteSetBstMidEndSepPunct{\mcitedefaultmidpunct}
{\mcitedefaultendpunct}{\mcitedefaultseppunct}\relax
\EndOfBibitem
\bibitem[Clark \emph{et~al.}(2007)Clark, Langeslag, Figdor, and van Leeuwen]{clark2007myosin}
K.~Clark, M.~Langeslag, C.~G. Figdor and F.~N. van Leeuwen, \emph{Trends in cell biology}, 2007, \textbf{17}, 178--186\relax
\mciteBstWouldAddEndPuncttrue
\mciteSetBstMidEndSepPunct{\mcitedefaultmidpunct}
{\mcitedefaultendpunct}{\mcitedefaultseppunct}\relax
\EndOfBibitem
\bibitem[Giomi \emph{et~al.}(2014)Giomi, Bowick, Mishra, Sknepnek, and Cristina~Marchetti]{giomi_defect_2014}
L.~Giomi, M.~J. Bowick, P.~Mishra, R.~Sknepnek and M.~Cristina~Marchetti, \emph{Philosophical Transactions of the Royal Society A: Mathematical, Physical and Engineering Sciences}, 2014, \textbf{372}, 20130365\relax
\mciteBstWouldAddEndPuncttrue
\mciteSetBstMidEndSepPunct{\mcitedefaultmidpunct}
{\mcitedefaultendpunct}{\mcitedefaultseppunct}\relax
\EndOfBibitem
\bibitem[Balasubramaniam \emph{et~al.}(2021)Balasubramaniam, Doostmohammadi, Saw, Narayana, Mueller, Dang, Thomas, Gupta, Sonam, Yap, Toyama, Mège, Yeomans, and Ladoux]{balasubramaniam_investigating_2021}
L.~Balasubramaniam, A.~Doostmohammadi, T.~B. Saw, G.~H. N.~S. Narayana, R.~Mueller, T.~Dang, M.~Thomas, S.~Gupta, S.~Sonam, A.~S. Yap, Y.~Toyama, R.-M. Mège, J.~M. Yeomans and B.~Ladoux, \emph{Nature Materials}, 2021,  1--11\relax
\mciteBstWouldAddEndPuncttrue
\mciteSetBstMidEndSepPunct{\mcitedefaultmidpunct}
{\mcitedefaultendpunct}{\mcitedefaultseppunct}\relax
\EndOfBibitem
\bibitem[Killeen \emph{et~al.}(2022)Killeen, Bertrand, and Lee]{killeen_polar_2022}
A.~Killeen, T.~Bertrand and C.~F. Lee, \emph{Physical Review Letters}, 2022, \textbf{128}, 078001\relax
\mciteBstWouldAddEndPuncttrue
\mciteSetBstMidEndSepPunct{\mcitedefaultmidpunct}
{\mcitedefaultendpunct}{\mcitedefaultseppunct}\relax
\EndOfBibitem
\bibitem[Shankar \emph{et~al.}(2022)Shankar, Souslov, Bowick, Marchetti, and Vitelli]{shankar2022topological}
S.~Shankar, A.~Souslov, M.~J. Bowick, M.~C. Marchetti and V.~Vitelli, \emph{Nature Reviews Physics}, 2022, \textbf{4}, 380--398\relax
\mciteBstWouldAddEndPuncttrue
\mciteSetBstMidEndSepPunct{\mcitedefaultmidpunct}
{\mcitedefaultendpunct}{\mcitedefaultseppunct}\relax
\EndOfBibitem
\bibitem[Plotnikov \emph{et~al.}(2012)Plotnikov, Pasapera, Sabass, and Waterman]{plotnikov_force_2012}
S.~Plotnikov, A.~Pasapera, B.~Sabass and C.~Waterman, \emph{Cell}, 2012, \textbf{151}, 1513--1527\relax
\mciteBstWouldAddEndPuncttrue
\mciteSetBstMidEndSepPunct{\mcitedefaultmidpunct}
{\mcitedefaultendpunct}{\mcitedefaultseppunct}\relax
\EndOfBibitem
\bibitem[Guolla \emph{et~al.}(2012)Guolla, Bertrand, Haase, and Pelling]{guolla_force_2012}
L.~Guolla, M.~Bertrand, K.~Haase and A.~E. Pelling, \emph{Journal of Cell Science}, 2012, \textbf{125}, 603--613\relax
\mciteBstWouldAddEndPuncttrue
\mciteSetBstMidEndSepPunct{\mcitedefaultmidpunct}
{\mcitedefaultendpunct}{\mcitedefaultseppunct}\relax
\EndOfBibitem
\bibitem[Zehnder \emph{et~al.}(2015)Zehnder, Suaris, Bellaire, and Angelini]{zehnder_cell_2015}
S.~Zehnder, M.~Suaris, M.~Bellaire and T.~Angelini, \emph{Biophysical Journal}, 2015, \textbf{108}, 247--250\relax
\mciteBstWouldAddEndPuncttrue
\mciteSetBstMidEndSepPunct{\mcitedefaultmidpunct}
{\mcitedefaultendpunct}{\mcitedefaultseppunct}\relax
\EndOfBibitem
\bibitem[Olenik \emph{et~al.}(2023)Olenik, Turley, Cross, Weavers, Martin, Chenchiah, and Liverpool]{olenik2023fluctuations}
M.~Olenik, J.~Turley, S.~Cross, H.~Weavers, P.~Martin, I.~V. Chenchiah and T.~B. Liverpool, \emph{Physical Review E}, 2023, \textbf{107}, 014403\relax
\mciteBstWouldAddEndPuncttrue
\mciteSetBstMidEndSepPunct{\mcitedefaultmidpunct}
{\mcitedefaultendpunct}{\mcitedefaultseppunct}\relax
\EndOfBibitem
\bibitem[Turlapati \emph{et~al.}(2017)Turlapati, Khan, Ramesh, Shamanna, Ghosh, and Rao]{turlapati2017elastic}
S.~Turlapati, R.~K. Khan, P.~Ramesh, J.~Shamanna, S.~Ghosh and N.~V. Rao, \emph{Liquid Crystals}, 2017, \textbf{44}, 784--797\relax
\mciteBstWouldAddEndPuncttrue
\mciteSetBstMidEndSepPunct{\mcitedefaultmidpunct}
{\mcitedefaultendpunct}{\mcitedefaultseppunct}\relax
\EndOfBibitem
\end{mcitethebibliography}
\bibliographystyle{rsc} 

\end{document}